\documentclass[aps,prl,reprint,superscriptaddress,longbibliography,floatfix]{revtex4-2}

\usepackage{amsmath}
\usepackage{amssymb}
\usepackage{graphicx}
\usepackage{bm}
\usepackage{hyperref}

\begin{document}

\title{Exact Law of Quantum Reversibility under Gaussian Pure Loss}

\author{Ammar Fayad}
\affiliation{Massachusetts Institute of Technology, Cambridge, Massachusetts 02139, USA}

\begin{abstract}
Classical reverse diffusion is generated by changing the drift at fixed noise. We show that the quantum version of this principle obeys an exact law with a sharp phase boundary. For Gaussian pure-loss dynamics---the canonical model of continuous-variable decoherence in optical attenuation channels, squeezed-light interferometric sensing, and superconducting bosonic architectures---complete positivity, the requirement that the dynamics remain physical even for systems entangled with an ancilla, creates an exact phase boundary at which the minimum reverse cost vanishes, fixes the reverse-noise budget on both sides, and makes pure nonclassical targets dynamically singular. The minimum reverse cost vanishes exactly at a critical squeezing-to-thermal ratio and is strictly positive away from it, with a sharp asymmetry: below the boundary, standard reverse prescriptions such as the fixed-diffusion Bayes reverse remain feasible at mild cost; above it, these prescriptions become infeasible, the covariance-aligned generator remains CP-feasible and uniquely optimal, and the cost can be severe. The optimal reverse noise is locked to the state's own fluctuation geometry and simultaneously minimizes the geometric, metrological, and thermodynamic price of reversal. For multimode trajectories, the exact cost is additive in a canonical set of mode-resolved data, and a globally continuous protocol attains this optimum on every mixed-state interval. If a pure nonclassical endpoint is included, the same pointwise law holds for every $t>0$, but the optimum diverges as $2/t$: exact reversal of a pure quantum state is dynamically unattainable. These results establish an exact law of quantum reversibility in the canonical pure-loss setting and provide a sharp benchmark for broader theories of quantum reverse diffusion.
\end{abstract}

\maketitle

Reversing an irreversible process without paying extra noise is one of the deepest idealizations behind classical diffusion theory~\cite{Anderson1982,Haussmann1986}. In modern score-based language, reverse diffusion is generated by changing the drift while keeping the noise level fixed~\cite{SohlDickstein2015,SongErmon2019,Ho2020,Song2021}. In quantum mechanics, however, complete positivity couples drift and diffusion at the generator level, so the classical reverse-by-drift picture cannot simply be transplanted~\cite{HHW2010,HolevoGiovannetti2012,gardiner2004quantum}. Recent work has made the quantum version of this problem newly urgent by identifying a semiclassical bridge between Bayes reverse diffusion and quantum reversibility via the Petz map~\cite{Petz1988,Kwon2022,Parzygnat2023,Nasu2025}. The natural question is whether quantum reverse dynamics obeys a comparably sharp law, or whether the extra constraints of quantum mechanics destroy any clean analogue of classical reversibility.

We show that the answer is an exact law. The unrestricted quantum reverse optimum exhibits a sharp phase boundary: the minimum reverse cost vanishes exactly at a critical squeezing-to-thermal ratio and is strictly positive on both sides, but with a sharp asymmetry---below the boundary, standard reverse prescriptions such as the fixed-diffusion Bayes reverse remain admissible at mild cost; above it, these prescriptions become infeasible, the covariance-aligned generator remains CP-feasible and uniquely optimal, and the cost can be severe. Here the phase boundary is a sharp nonanalytic change in the active complete-positivity constraints of the Gaussian reverse optimization problem. The optimal reverse protocol simultaneously minimizes three independently motivated physical quantities---the displacement quantum Fisher-information injection rate, the displacement-Bures rate, and the fluctuation entropy production---so the boundary at which all three costs vanish together is the unique point of free reversal. Pure nonclassical states are singular: the exact pointwise optimum diverges universally as $2/t$, so no finite continuous protocol can reach a pure quantum endpoint.

We establish this exactly in the canonical pure-loss continuous-variable setting~\cite{Braunstein2005,Weedbrook2012,HHW2010}, within the class of Gaussian Markov generators. A Gaussian Markov generator $(K_t,D_t)$ evolves the covariance matrix $\Gamma_t$ by
\begin{equation}
\dot{\Gamma}_t = K_t\Gamma_t + \Gamma_t K_t^T + D_t,
\label{eq:covariance_evolution}
\end{equation}
and complete positivity requires
\begin{equation}
D_t + i\!\left(K_t\sigma + \sigma K_t^T\right) \succeq 0.
\label{eq:cp_condition}
\end{equation}
This is the decisive quantum departure from the classical reverse-diffusion picture. In classical Fokker--Planck theory, the drift field can be changed independently of the diffusion tensor, so reversing a diffusion amounts to changing the drift while leaving the noise level fixed. Eq.~(\ref{eq:cp_condition}) shows that quantum mechanics forbids this independence at the generator level: complete positivity imposes a joint constraint on drift and diffusion through the canonical commutation structure. Reducing reverse noise below the level demanded by Eq.~(\ref{eq:cp_condition}) is physically forbidden, not just suboptimal. Complete positivity means that the reverse dynamics must send every valid quantum state---including states entangled with an external ancilla---to another valid quantum state. Violating Eq.~(\ref{eq:cp_condition}) would produce a nonphysical output. 

For one mode, the forward pure-loss dynamics is
\begin{equation}
K_{\mathrm{fwd}} = -\gamma I_2, \qquad D_{\mathrm{fwd}} = 2\gamma I_2,
\label{eq:forward_pure_loss_generator}
\end{equation}
with covariance path
\begin{equation}
\Gamma_t = e^{-2\gamma t}\Gamma_0 + (1 - e^{-2\gamma t})I_2.
\label{eq:forward_pure_loss_path}
\end{equation}
This describes photon loss to the environment at rate $\gamma$: the covariance matrix relaxes toward the vacuum value $I_2$ as excitations leak out of the mode. Pure loss is the standard effective model for optical attenuation and bosonic energy relaxation, so the same covariance law underlies lossy fiber links, squeezed-light preparation and propagation chains, and bosonic cavity architectures. In squeezed-light interferometric sensing---including gravitational-wave metrology---the full device is a much richer frequency-dependent open system, but optical loss remains the basic mechanism that degrades injected squeezing, making pure loss the canonical benchmark against which more complicated sensing chains are judged. By time $t$, the excess covariance above vacuum has been suppressed by the factor $e^{-2\gamma t}$. For a squeezed-thermal target
\begin{equation}
\Gamma_0 = \operatorname{diag}(\nu e^{2r}, \nu e^{-2r}),
\label{eq:squeezed_thermal_target}
\end{equation}
one can construct a fixed-diffusion Bayes reverse candidate that keeps the forward noise and modifies only the drift. This candidate is completely positive if and only if $\cosh(2r) \le \nu$~\cite{fayad2026quantum}. Our first main result is that this boundary is not an artifact of the Bayes ansatz: it is the exact point at which the unrestricted instantaneous Gaussian reverse cost vanishes. Away from this boundary the cost is strictly positive, but the two sides are qualitatively different. Below it, the Bayes reverse and other standard prescriptions remain feasible, and the cost is controlled by the mild denominator $\nu+1$. Above it, standard reverse prescriptions such as the fixed-diffusion Bayes reverse and the best isotropic protocol with $D=cI_2$ become infeasible, the cost is controlled by the denominator $\nu-1$, and the covariance-aligned generator remains CP-feasible and uniquely optimal. Physically, this is because preserving a strongly squeezed quadrature during reversal requires an increasingly aggressive anti-damping drift along the fragile direction; complete positivity then forces a corresponding increase in diffusion, so a fixed-diffusion ansatz---and even the best isotropic repair---can no longer remain physical once $\cosh(2r)>\nu$.

The cost functional that the optimum minimizes,
\begin{equation}
Z(D;\Gamma) := \operatorname{Tr}(\Gamma^{-1}D),
\label{eq:cost_functional}
\end{equation}
is the natural reverse-diffusion budget: it measures the rate at which reverse noise is injected, weighted to penalize noise along the state's most fragile low-variance quadrature directions---precisely the directions where nonclassical squeezing is most easily degraded. Operationally, a cost of $Z=10\gamma$ means that the reverse protocol must inject noise at a rate ten times the forward pure-loss scale, weighted by the state's own fluctuation geometry rather than counted isotropically. The same quantity admits three independent physical interpretations. For displacement estimation of a Gaussian state, the symmetric-logarithmic-derivative quantum Fisher matrix is $\Gamma^{-1}$, so diffusion $D$ injects displacement Fisher information at the rate $\operatorname{Tr}(\Gamma^{-1}D)$~\cite{Pinel2013GaussianQFI,Liu2020QFIM}. Since the quantum Fisher metric is four times the Bures metric, the same quantity is four times the displacement-Bures rate~\cite{Petz1996,Pinel2013GaussianQFI,Liu2020QFIM}. The Gaussian de Bruijn identity decomposes entropy production under a Gaussian semigroup into a drift part and a fluctuation part proportional to $\operatorname{Tr}(\Gamma^{-1}D)$, so the same optimizer also minimizes the fluctuation contribution to entropy production~\cite{Toscano2021}. The same boundary and the same optimizer therefore govern the geometric, metrological, and thermodynamic price of reversal.

The one-mode law admits a full multimode completion. For multimode pure-loss trajectories, the exact reverse cost remains additive in a canonical set of mode-resolved anti-squeezing data, and this additive law is the exact unrestricted Gaussian optimum on every interval for which all symplectic eigenvalues remain strictly above one. As proved in the Supplemental Material by constructing a globally continuous moving Williamson frame from an analytic eigenbasis of $i\Gamma_t^{1/2}\sigma\Gamma_t^{1/2}$, crossings and degeneracies do not destroy exactness: a globally continuous moving Williamson frame reduces the multimode problem to independent scalar mode optimizations. If a pure nonclassical endpoint is included, the same pointwise law holds for every $t > 0$, but the optimum diverges universally as $2/t$. Exact Gaussian reversal of a pure nonclassical target is therefore dynamically unattainable.

The result is relevant beyond Gaussian channel theory. Reverse-diffusion constructions are now discussed in quantum settings partly through the Bayes--Petz correspondence~\cite{Petz1988,Kwon2022,Parzygnat2023,Nasu2025}, and our law identifies the unique zero-cost boundary of the unrestricted Gaussian reverse problem, the asymmetric positive-cost regimes on either side, and the feasibility wall beyond which fixed-diffusion reverse constructions become unphysical.  In the same canonical one-mode pure-loss setting, the continuous-time Gaussian Petz reverse is strictly suboptimal for every squeezed target, so the present result supplies not only a feasibility boundary but the exact Gaussian optimizer that replaces the most natural Bayes/Petz reverse construction. As shown in Fig.~\ref{fig:1}(a), published $1550\,\mathrm{nm}$ squeezing demonstrations~\cite{Vahlbruch2016,Mehmet2011,Meylahn2022} already occupy the part of state space where the reversibility law is operationally active: current high-squeezing sources lie well inside the reverse-noise-forced sector $\nu<\cosh(2r)$, where standard fixed-diffusion reverse prescriptions are infeasible and the exact optimum must inject reverse diffusion. This is directly relevant to squeezed-light platforms in which optical loss is the dominant mechanism that degrades nonclassical noise suppression. Recent full-scale gravitational-wave detectors now operate with frequency-dependent squeezed-light enhancement and have even pushed quantum noise below the standard quantum limit in part of the detection band~\cite{Ganapathy2023LIGOFD,Jia2024SQL}. The present theorem is not a model of the full interferometer; rather, it identifies the pure-loss benchmark that injected squeezed states face before and within more complicated sensing architectures.

For Gaussian pure-loss dynamics, reversibility is governed by an exact cost law: the minimum reverse cost vanishes at $\cosh(2r) = \nu$, is strictly positive on both sides with qualitatively different regimes, and diverges at pure nonclassical endpoints.

\paragraph*{Bayes reverse threshold---}
For the one-mode pure-loss channel, the fixed-diffusion Bayes reverse candidate keeps the forward diffusion and modifies only the drift,
\begin{equation}
K^{\mathrm{Bayes}}=K_{\mathrm{fwd}}+D_{\mathrm{fwd}}\Gamma_0^{-1},
\qquad
D^{\mathrm{Bayes}}=D_{\mathrm{fwd}}.
\label{eq:bayes_generator}
\end{equation}
Its complete-positivity matrix is
\begin{equation}
M^{\mathrm{Bayes}} = D^{\mathrm{Bayes}} + i\!\left(K^{\mathrm{Bayes}}\sigma+\sigma(K^{\mathrm{Bayes}})^T\right).
\label{eq:bayes_cp_matrix}
\end{equation}
For the $i\sigma$-eigenvector $u=\frac{1}{\sqrt2}(1,i)^T$, one finds
\begin{equation}
u^\dagger M^{\mathrm{Bayes}}u = 4\gamma\left(1-\frac{\cosh(2r)}{\nu}\right).
\label{eq:bayes_cp_expectation}
\end{equation}
Hence the fixed-diffusion Bayes reverse lift is completely positive if and only if
\begin{equation}
\cosh(2r)\le \nu.
\label{eq:bayes_threshold}
\end{equation}
A squeezing level of about $13\,\mathrm{dB}$ in quadrature variance corresponds to $r\approx 1.5$, for which $\cosh(2r)=\cosh(3)\approx 10.1$. Thus the fixed-diffusion Bayes reverse remains physical only if the thermal parameter is at least $\nu\approx 10.1$, equivalently $\bar n=(\nu-1)/2\gtrsim 4.5$ thermal photons. Strong squeezing therefore drives the Bayes reverse out of the CP-feasible region unless the state is correspondingly thermalized.

\begin{figure}[b]
\centering
\includegraphics[width=\linewidth]{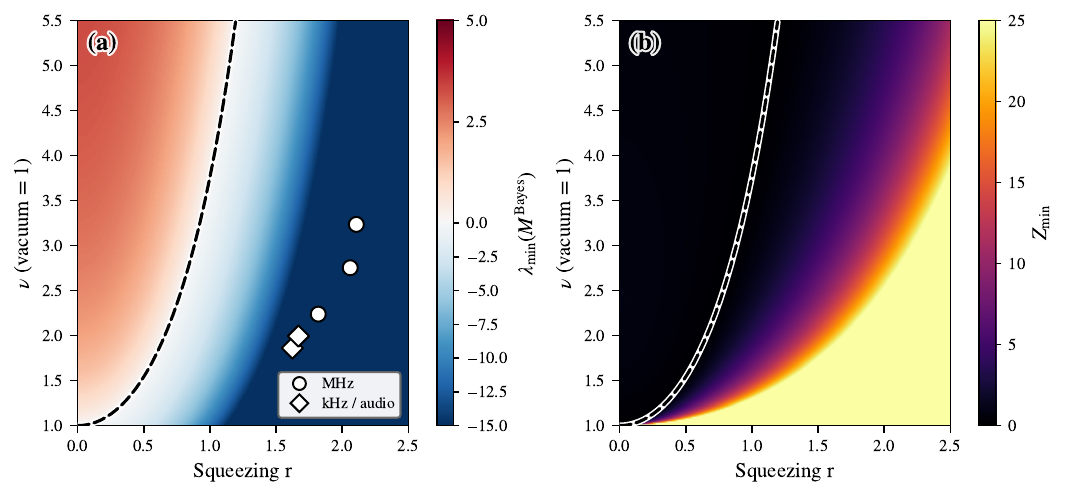}
\caption{Sharp phase boundary for Gaussian quantum reversibility under pure loss. (a) Minimum eigenvalue of the fixed-diffusion Bayes reverse CP matrix $M^{\mathrm{Bayes}}$ versus squeezing $r$ and thermal variance $\nu$ (vacuum $=1$). The dashed curve is the exact threshold $\nu=\cosh(2r)$. Published $1550\,\mathrm{nm}$ squeezing demonstrations~\cite{Mehmet2011,Meylahn2022} are overlaid by mapping directly reported squeezed and anti-squeezed quadrature variances to $(r,\nu)$; all displayed operating points lie in the sector $\nu<\cosh(2r)$, locating current hardware in the reverse-noise-forced region of the phase diagram. (b) Exact optimal reverse cost $Z_{\min}$ for the unrestricted instantaneous one-mode Gaussian reverse problem. The cost vanishes exactly on the dashed curve and is strictly positive on both sides, with qualitatively different regimes above and below.}
\label{fig:1}
\end{figure}

Figure~\ref{fig:1} shows the resulting phase boundary. Beyond it, the direct classical reverse-drift prescription becomes unphysical and the optimal reverse cost enters the severe branch controlled by $\nu-1$. Below it, the cost is also strictly positive but mild, controlled by $\nu+1$, and standard reverse prescriptions remain admissible.

\paragraph*{Exact instantaneous one-mode optimum---}
We now determine the minimum noise that must be injected to reverse pure-loss decoherence of a given quantum state, optimizing over all Gaussian reverse generators consistent with complete positivity. Among all one-mode Gaussian generators $(K,D)$ satisfying exact reverse covariance matching at $\Gamma_0$ and generator complete positivity, we minimize $Z(D;\Gamma_0)$. Writing
\begin{equation}
x:=\frac{\cosh(2r)}{\nu},
\label{eq:def_x}
\end{equation}
the exact optimum is
\begin{equation}
Z_{\min} = \frac{4\gamma|x-1|}{\nu-\operatorname{sgn}(x-1)}.
\label{eq:zmin}
\end{equation}
As shown in the Supplemental Material via an explicit semidefinite-program duality argument, exact rank-one dual witnesses certify Eq.~(\ref{eq:zmin}), and complementary slackness forces covariance alignment. Equivalently,
\begin{equation}
x<1 \Rightarrow Z_{\min}=\frac{4\gamma(1-x)}{\nu+1},
\qquad
x>1 \Rightarrow Z_{\min}=\frac{4\gamma(x-1)}{\nu-1}.
\label{eq:zmin_branches}
\end{equation}
The cost vanishes at $x=1$ and is strictly positive on both sides, but with a sharp asymmetry:
the above-threshold denominator $\nu-1$ can be much smaller than the below-threshold denominator $\nu+1$, making the cost above threshold potentially far more severe. The formula passes two immediate physical sanity checks. If $r=0$ then $x=1/\nu<1$ and
\begin{equation}
Z_{\min}=\frac{4\gamma(1-1/\nu)}{\nu+1}
=\frac{4\gamma(\nu-1)}{\nu(\nu+1)}
\sim \frac{4\gamma}{\nu}\qquad (\nu\gg 1),
\label{eq:thermal_limit}
\end{equation}
so highly thermal states become easy to reverse: as thermal variance grows, the required reverse cost per unit time falls to zero. If $\nu=1$ and $r>0$, then $x=\cosh(2r)>1$ and Eq.~(\ref{eq:zmin}) forces divergence through the denominator $\nu-1$. The exact formula itself already signals that pure squeezed targets are singular, anticipating the pure-endpoint law derived below. Moreover, the unique optimal diffusion tensor is covariance-aligned:
\begin{equation}
D_{\mathrm{opt}}=\frac{Z_{\min}}{2}\Gamma_0.
\label{eq:dopt}
\end{equation}
The optimal reverse noise is locked to the state's own fluctuation geometry---a structural consequence of the KKT conditions, not an ansatz. Concretely, the optimal reverse protocol injects diffusion in proportion to the state's own quadrature variances: more along the anti-squeezed direction and less along the squeezed direction, matching the elliptical shape of the state's fluctuation profile. This is a fundamental Gaussian bound, not a near-term engineering prescription, since dynamically realizing a reservoir whose covariance and principal axes track the state's fluctuation geometry would itself be a demanding control task. The exact noiseless transition of the unrestricted Gaussian reverse problem is therefore
\begin{equation}
Z_{\min}=0 \iff \cosh(2r)=\nu.
\label{eq:zero_cost_boundary}
\end{equation}
The derivation is structurally rigid: exact rank-one dual witnesses certify the optimum, and complementary slackness forces covariance alignment rather than assuming it. The boundary at $\cosh(2r)=\nu$ is therefore not a property of one failed reverse ansatz; it is the exact zero of the unrestricted cost function, flanked by qualitatively distinct nonzero-cost regimes.

\paragraph*{Physical currency of reverse diffusion---}
The cost $Z(D;\Gamma)$ is the physically natural reverse budget. For faithful one-mode targets, displacement estimation provides the first interpretation: if the state is displaced in phase space by a small parameter vector, then the symmetric-logarithmic-derivative quantum Fisher matrix is $\Gamma^{-1}$, so a diffusion increment $D\,dt$ injects displacement Fisher information at the rate $\operatorname{Tr}(\Gamma^{-1}D)$~\cite{Pinel2013GaussianQFI,Liu2020QFIM}. The second interpretation is geometric: because the quantum Fisher metric is four times the Bures metric, the same quantity is exactly four times the displacement-Bures rate~\cite{Petz1996,Pinel2013GaussianQFI,Liu2020QFIM}. The third is thermodynamic: the Gaussian de Bruijn identity separates entropy production into a drift contribution and a fluctuation contribution, and the latter is proportional to $\operatorname{Tr}(\Gamma^{-1}D)$~\cite{Toscano2021}. The optimizer found here therefore minimizes, with the same generator and at the same boundary, the metrological information injected by reverse noise, the geometric distance traversed in state space, and the fluctuation part of the entropy-production budget. At the threshold $\cosh(2r)=\nu$, all three vanish together---the unique point at which reversal is free in every physical currency simultaneously.

\paragraph*{Multimode exact law---}
For multimode pure-loss trajectories, the reverse cost obeys a gauge-invariant additive lower bound
\begin{equation}
Z(D_t;\Gamma_t)\ge \sum_k \frac{4\gamma|x_k^*(t)-1|}{\nu_k(t)-\operatorname{sgn}(x_k^*(t)-1)}.
\label{eq:lower_bound}
\end{equation}
Here $x_k^*(t)$ are the canonical anti-squeezing data defined by the moving Williamson frame introduced below. Eq.~(\ref{eq:lower_bound}) says that dangerous modes contribute additively: neither spectator modes nor intermode structure can hide the reverse cost. For familiar entangled Gaussian resources such as two-mode squeezed states used in continuous-variable teleportation, entanglement distribution, and continuous-variable key distribution~\cite{Braunstein2005,Weedbrook2012}, the message is concrete: under pure loss, the exact Gaussian reverse cost is still the sum of canonical mode-resolved contributions in the moving Williamson frame. Entanglement does not create a collective shortcut around the reverse-noise budget.

The lower bound is also exact. The multimode pure-loss path admits a globally continuous moving Williamson frame
\begin{equation}
\Gamma_t=S_c(t)\Lambda_t S_c(t)^T, \qquad \Lambda_t=\bigoplus_{k=1}^N \nu_k(t)I_2,
\label{eq:moving_williamson_frame}
\end{equation}
for which the reverse problem obeys a simple source equation. Defining
\begin{equation}
W_c:=S_c^{-1}\dot S_c,\qquad G_t:=S_c^{-1}S_c^{-T},
\label{eq:wc_gt_defs}
\end{equation}
one finds
\begin{equation}
R_t:=-\dot{\Lambda}_t = 2\gamma(\Lambda_t-G_t)+W_c\Lambda_t+\Lambda_t W_c^T.
\label{eq:source_equation_real}
\end{equation}
In the complex canonical representation this becomes
\begin{equation}
R_t^C = 2\gamma(\Lambda_t^C-G_t^C)+[W_t^C,\Lambda_t^C].
\label{eq:source_equation_complex}
\end{equation}
Its off-diagonal entries give
\begin{equation}
(\nu_k-\nu_j)W_{t,jk}^C=2\gamma G_{t,jk}^C, \qquad j\neq k,
\label{eq:offdiag_entries}
\end{equation}
while the diagonal entries give the exact scalar reverse sources
\begin{equation}
s_k(t)=2\gamma\nu_k(t)\bigl(1-x_k^*(t)\bigr).
\label{eq:scalar_reverse_sources}
\end{equation}
Crossings are therefore non-singular: the analytic eigenbasis of $i\Gamma_t^{1/2}\sigma\Gamma_t^{1/2}$ obeys a bilinear orthogonality identity that eliminates the dangerous off-diagonal metric couplings at degeneracies (see Supplemental Material), reducing the multimode problem to independent one-mode scalar optimizations. The lower bound in Eq.~(\ref{eq:lower_bound}) is attained by a globally continuous Gaussian CP reverse protocol, giving the exact unrestricted Gaussian optimum on every mixed-state interval.

\paragraph*{Pure-endpoint singularity---}
A pure squeezed state lies on the extreme boundary of the physically allowed Gaussian state space, with no mixedness buffer. Reversing decoherence all the way back to such a state is therefore singular in cost: if a pure nonclassical endpoint is included, the same pointwise law remains valid for every $t>0$, but the optimum diverges universally as $2/t$, as derived asymptotically in the Supplemental Material:
\begin{equation}
Z_{\min}(t)=\frac{2}{t}+O(1), \qquad t\to 0^+.
\label{eq:pure_endpoint_asymptotic}
\end{equation}
The coefficient is independent of the squeezing parameter $r$: once the target lies on the pure nonclassical boundary, the leading reverse cost no longer remembers how strongly squeezed the state is, only that one is attempting to reach a pure one-mode Gaussian endpoint. The coefficient $2$ is naturally associated with the two canonical quadrature directions of a single mode; the singularity is structural rather than state-specific. This is an irreversibility statement within the continuous Gaussian Markov reverse class. Exact reversal to a pure nonclassical target demands unbounded pointwise reverse cost, even though the target is reached only at a single endpoint in time.

The integrated cost is milder than the pointwise divergence might suggest. Since $Z_{\min}(t)\sim 2/t$, the accumulated reverse action over $[\varepsilon,T]$ diverges only logarithmically,
\begin{equation}
\int_{\varepsilon}^{T} Z_{\min}(t)\,dt = 2\ln\!\frac{T}{\varepsilon}+O(1)
\qquad (\varepsilon\to 0^+).
\label{eq:integrated_reverse_action}
\end{equation}
Exact pure-state reversal is therefore unattainable in this class, but approximate reversal toward very high purity remains quantitatively meaningful: the obstruction is singular, yet only logarithmically so at the level of total action. This is a second-law-like statement for the continuous Gaussian reverse problem: exact reversal of a pure nonclassical state requires unbounded thermodynamic fuel, linking the present reversibility law to broader fluctuation-theoretic notions of irreversibility for quantum channels~\cite{KwonKim2019QFT}.

Figure~\ref{fig:2} shows this irreversibility explicitly: the pointwise optimum approaches the universal asymptote $2/t$, while the rescaled quantity $t Z_{\min}(t)$ collapses to the universal coefficient $2$, independent of squeezing. The figure isolates a universal one-mode boundary law: pure nonclassical Gaussian endpoints are dynamically inaccessible to any finite continuous Gaussian reverse protocol.

\begin{figure}[t]
\centering
\includegraphics[width=\linewidth]{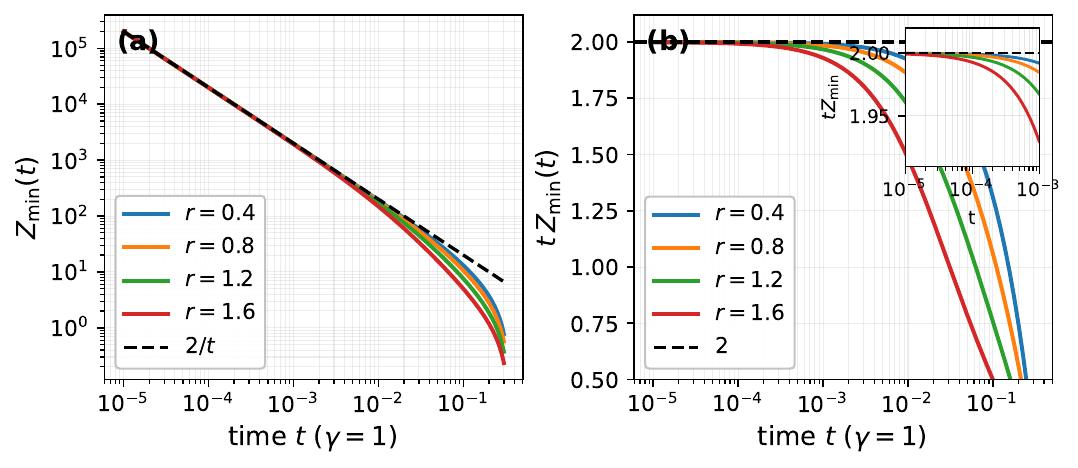}
\caption{Pure-endpoint singularity of the exact Gaussian reversibility law. For pure squeezed targets, the exact pointwise Gaussian reverse optimum diverges as $Z_{\min}(t)\sim 2/t$ as $t\to 0^+$. (a) The optimal cost approaches the universal asymptote $2/t$. (b) The rescaled quantity $t Z_{\min}(t)$ collapses to the universal coefficient $2$, independent of the squeezing parameter $r$.}
\label{fig:2}
\end{figure}

\begin{figure}[h]
\centering
\includegraphics[width=\linewidth]{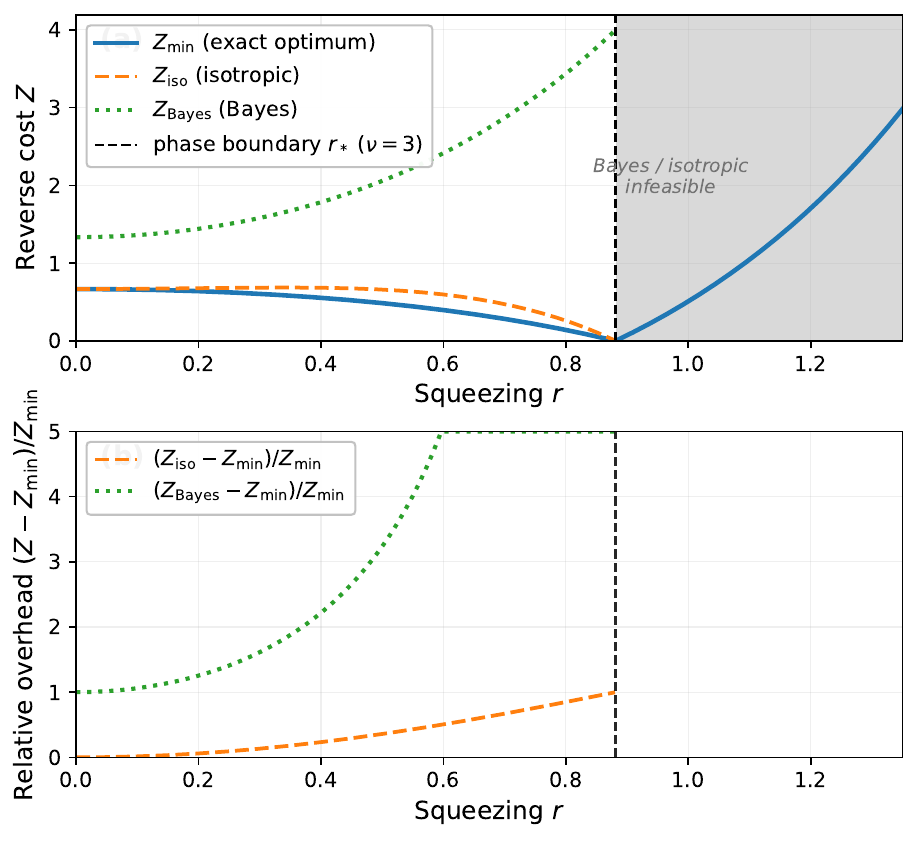}
\caption{Reverse-cost comparison of three Gaussian reverse protocols for a squeezed-thermal target under pure loss ($\nu=3$). The exact optimum $Z_{\min}$ (solid), the best isotropic protocol with $D=cI_2$ (dashed), and the fixed-diffusion Bayes reverse (dotted). The vertical dashed line marks the exact phase boundary $\cosh(2r)=\nu$, where $Z_{\min}$ vanishes. Below threshold, all three protocols are feasible but naive prescriptions pay a strict penalty relative to the exact optimum; above threshold, both Bayes and isotropic protocols become infeasible while the covariance-aligned generator remains CP-feasible and uniquely optimal---the difference is not a formal refinement but the difference between feasible and impossible reversal for standard prescriptions in experimentally relevant regimes. Above threshold, isotropic repair does not eventually restore complete positivity because exact covariance matching simultaneously drives the reverse drift, so the symplectic term $i(K\sigma+\sigma K^T)$ grows with the injected isotropic noise and overwhelms the diagonal repair in the most squeezed direction.}
\label{fig:3}
\end{figure}

\paragraph*{Discussion}---We have derived an exact law of Gaussian quantum reversibility for pure-loss dynamics. The unrestricted instantaneous cost vanishes exactly at $\cosh(2r)=\nu$ and is strictly positive on both sides, with qualitatively different cost regimes and protocol feasibility above and below the boundary. The multimode optimum is exactly additive on every mixed-state interval, and pure nonclassical endpoints are singular. The same cost functional controls the geometric, metrological, and thermodynamic fuel required for reversal, so the law identifies the exact price of reversing an irreversible quantum process within the continuous Gaussian Markovian class.


The result also sharpens the Bayes--Petz reverse-diffusion program~\cite{Petz1988,Kwon2022,Parzygnat2023,Nasu2025}. In the canonical one-mode pure-loss setting, the fixed-diffusion reverse prescription that works classically encounters the exact complete-positivity boundary $\cosh(2r)=\nu$. If $Z_{\mathrm{Petz}}$ denotes the covariance-weighted diffusion cost of the continuous-time Gaussian Petz reverse, then
\begin{equation}
Z_{\mathrm{Petz}}-Z_{\min}=
\begin{cases}
\dfrac{4\gamma(\cosh(2r)-1)}{\nu-1}, & \cosh(2r)\le \nu,\\[6pt]
\dfrac{4\gamma(\cosh(2r)+1)}{\nu+1}, & \cosh(2r)>\nu,
\end{cases}
\label{eq:petz_gap}
\end{equation}
which is strictly positive for every squeezed target $r>0$, with equality only for unsqueezed thermal states. The present work therefore identifies the exact unrestricted Gaussian optimum that improves on the natural Bayes/Petz lift in the first canonical setting where the reverse problem can be solved exactly.
In practical terms, the law answers a concrete question faced by any laboratory attempting to undo pure-loss decoherence of a nonclassical state within the continuous Gaussian setting: what is the minimum reverse noise that must be injected?
The answer is given exactly by Eq.~(\ref{eq:zmin}); it depends only on the state's squeezing and thermal parameters and diverges for pure states---setting a fundamental floor on the Gaussian noise cost of quantum state recovery. In the canonical pure-loss setting, a classical reverse-drift picture does not simply fail heuristically; it is replaced by a sharp reversibility law with an exact optimizer, a unique zero-cost boundary, and a singular pure endpoint. To place the law against current hardware, we overlay published $1550\,\mathrm{nm}$ squeezing demonstrations on Fig.~\ref{fig:1}(a), mapping reported squeezed and anti-squeezed quadrature variances $(V_-,V_+)$ to the theorem variables via $\nu=\sqrt{V_-V_+}$ and $r=\tfrac14\ln(V_+/V_-)$. The directly reported pairs $(12.3/19.3)\,\mathrm{dB}$ and $(11.4/16.8)\,\mathrm{dB}$ from Mehmet \emph{et al.}~\cite{Mehmet2011} and $(13.5/22.3)\,\mathrm{dB}$, $(13.2/23.4)\,\mathrm{dB}$, and $(11.5/17.5)\,\mathrm{dB}$ from Meylahn \emph{et al.}~\cite{Meylahn2022} all map into the sector $\nu<\cosh(2r)$. Experimentally realized squeezing regimes therefore already occupy the part of state space where the reversibility law is operationally active. Figure~\ref{fig:3} makes the operational consequence explicit: the exact optimum, the fixed-diffusion Bayes reverse, and the best isotropic protocol are compared for $\nu=3$, and the difference between the exact law and naive prescriptions is the difference between feasible and impossible reversal. For highly nonclassical bosonic targets, the pure-endpoint divergence identifies an intrinsic obstruction to exact Gaussian reversal that precedes any specific decoder architecture. For quantum diffusion models aiming to prepare or denoise continuous-variable states, the law gives a concrete feasibility statement: a reverse construction that keeps the diffusion fixed and changes only the drift is admissible only on one side of the boundary; beyond it, such a prescription is unphysical. Complete positivity enforces an exact, state-dependent reverse-noise budget and a singular pure-state boundary already in the canonical setting.

A neighboring community where the result has direct quantitative consequences is bosonic quantum error correction, especially finite-energy Gottesman--Kitaev--Preskill (GKP) encodings in superconducting cavities~\cite{Sivak2023BreakEven,LachanceQuirion2024AutonomousGKP,Brady2024GKPReview}. Practical GKP recovery relies on syndrome measurements, ancilla couplings, feedforward, reservoir engineering, and other explicitly non-Gaussian resources, so the present results do not constrain those architectures directly. What they do constrain is the covariance-level performance of any recovery strategy confined to exact continuous Gaussian semigroups. For an isotropic covariance target $\Gamma_0=\nu I_2$, which provides the natural covariance benchmark associated with the second-moment sector of a square-lattice finite-energy GKP state, the exact one-mode optimum gives
\begin{equation}
D_{\mathrm{opt}}=\frac{Z_{\min}}{2}\Gamma_0
=\frac{2\gamma(\nu-1)}{\nu+1}I_2
=\frac{2\gamma\,\bar n}{\bar n+1}I_2,
\qquad
\bar n=\frac{\nu-1}{2}.
\end{equation}
Even the best exact continuous Gaussian reverse must therefore inject an absolute displacement-diffusion rate that approaches the forward pure-loss scale $2\gamma I_2$ as the code quality increases ($\bar n\gg1$). In the standard effective-displacement picture often used to analyze bosonic-code noise, the resulting round-trip displacement variance is amplified by the factor $(2\bar n+1)/(\bar n+1)$, approaching $2$ at large $\bar n$. Gaussian reversal is not a route around the bosonic error budget: beating this covariance-sector floor requires the non-Gaussian ingredients used by actual GKP architectures. Indeed, the leading superconducting GKP experiments implement recovery through measurement-based or autonomously stabilized non-Gaussian protocols, not continuous Gaussian semigroups.

The subsequent open problem is whether unrestricted non-Gaussian reverse protocols obey the same law, or whether they can beat the Gaussian optimum by using resources unavailable to any continuous Gaussian semigroup~\cite{BarnumKnill2002,FawziRenner2015,Junge2018}. The escape routes include syndrome-resolved measurements, ancilla-assisted feedback, and engineered non-Gaussian reservoirs---precisely the ingredients used in leading bosonic recovery architectures, so the question is experimentally meaningful. At the same time, the recurring role of Gaussian extremality in bosonic channel theory suggests that the exact optimizer found here may remain the correct benchmark even beyond the Gaussian class~\cite{Giovannetti2015GaussianOptimizer}. The present work establishes the exact law in the canonical pure-loss setting that provides a sharp benchmark for any broader theory of quantum reverse diffusion.
\nocite{rellich1969perturbation}

\bibliographystyle{apsrev4-2}
\bibliography{ref}

\clearpage
\onecolumngrid 

\begin{center}
  \textbf{\large Supplementary Material}
\end{center}
\vspace{1cm}

\setcounter{equation}{0}
\setcounter{figure}{0}
\setcounter{table}{0}
\setcounter{page}{1}
\setcounter{section}{0}

\renewcommand{\theequation}{S\arabic{equation}}
\renewcommand{\thefigure}{S\arabic{figure}}
\renewcommand{\thetable}{S\arabic{table}}
\renewcommand{\thesection}{S\arabic{section}}

\title{Supplemental Material}


\maketitle

\setcounter{equation}{0}
\renewcommand{\theequation}{S\arabic{equation}}
\setcounter{figure}{0}
\renewcommand{\thefigure}{S\arabic{figure}}
\setcounter{section}{0}
\renewcommand{\thesection}{S\arabic{section}}

\section{Conventions and generator-level complete positivity}

We use standard continuous-variable Gaussian conventions. The symplectic form is
\begin{equation}
\sigma=\bigoplus_{j=1}^N
\begin{pmatrix}
0&1\\
-1&0
\end{pmatrix}.
\end{equation}
A centered Gaussian channel with matrices $(X,Y)$ acts on covariance matrices by
\begin{equation}
\Gamma\mapsto X\Gamma X^T+Y.
\end{equation}
The Heinosaari--Holevo--Wolf complete-positivity condition~\cite{HHW2010} is
\begin{equation}
Y\succeq i(\sigma-X\sigma X^T).
\end{equation}
For an infinitesimal step,
\begin{equation}
X=I+K\,dt+o(dt),\qquad Y=D\,dt+o(dt),
\end{equation}
so
\begin{equation}
X\sigma X^T=\sigma+(K\sigma+\sigma K^T)dt+o(dt).
\end{equation}
Dividing by $dt$ gives the generator-level condition
\begin{equation}
M:=D+i(K\sigma+\sigma K^T)\succeq 0.
\label{eq:CP}
\end{equation}
This is the basic CP constraint used throughout.

\section{One-mode fixed-diffusion Bayes threshold}

For one-mode pure loss,
\begin{equation}
K_{\mathrm{fwd}}=-\gamma I_2,\qquad D_{\mathrm{fwd}}=2\gamma I_2.
\end{equation}
The fixed-diffusion Gaussian Bayes reverse lift keeps the forward diffusion and changes only the drift:
\begin{equation}
K_{\mathrm{Bayes}}=-\gamma I_2+2\gamma\Gamma^{-1},\qquad D_{\mathrm{Bayes}}=2\gamma I_2.
\end{equation}
For a squeezed-thermal target
\begin{equation}
\Gamma=\operatorname{diag}(\nu e^{2r},\nu e^{-2r}),
\end{equation}
the CP matrix is
\begin{equation}
M_{\mathrm{Bayes}}
=2\gamma(I_2-i\sigma)+2i\gamma(\Gamma^{-1}\sigma+\sigma\Gamma^{-1}).
\end{equation}
Let
\begin{equation}
u = \frac{1}{\sqrt 2}(1,i)^T,
\qquad \sigma u = i u.
\end{equation}
Then
\begin{equation}
u^\dagger M_{\mathrm{Bayes}}u
=4\gamma\left(1-u^\dagger\Gamma^{-1}u\right)
=4\gamma\left(1-\frac{\cosh(2r)}{\nu}\right).
\end{equation}
Because $\Gamma^{-1}$ is diagonal, $M_{\mathrm{Bayes}}$ is a real linear combination of $I_2$ and $i\sigma$, hence diagonal in the basis $\{u,u^*\}$. The second eigenvalue is $u^{*\dagger} M_{\mathrm{Bayes}} u^* = 4\gamma(1+\cosh(2r)/\nu) > 0$, so
\begin{equation}
M_{\mathrm{Bayes}}\succeq 0
\iff
\cosh(2r)\le \nu.
\end{equation}

\section{Exact unrestricted one-mode local problem as an SDP}

Fix the one-mode target covariance
\begin{equation}
\Gamma_0=\operatorname{diag}(v_q,v_p)=\operatorname{diag}(\nu e^{2r},\nu e^{-2r}).
\end{equation}
We seek a Gaussian reverse generator $(K,D)$ that matches the reverse covariance equation exactly at $\Gamma_0$:
\begin{equation}
K\Gamma_0+\Gamma_0K^T+D=2\gamma(\Gamma_0-I).
\label{eq:matching}
\end{equation}
The objective is the covariance-weighted diffusion cost
\begin{equation}
Z(D;\Gamma_0)=\operatorname{Tr}(\Gamma_0^{-1}D)=\frac{D_{qq}}{v_q}+\frac{D_{pp}}{v_p}.
\end{equation}
Write
\begin{equation}
D=
\begin{pmatrix}
a & c\\
c & b
\end{pmatrix},
\qquad
x:=\frac{\cosh(2r)}{\nu}.
\end{equation}
The diagonal drift entries are fixed by exact matching:
\begin{equation}
K_{qq}=\gamma\!\left(1-\frac{1}{v_q}\right)-\frac{a}{2v_q},
\qquad
K_{pp}=\gamma\!\left(1-\frac{1}{v_p}\right)-\frac{b}{2v_p}.
\end{equation}
Using the identity
\begin{equation}
A\sigma+\sigma A^T=(\operatorname{Tr}A)\,\sigma
\label{eq:trace_identity}
\end{equation}
valid for every real $2\times 2$ matrix $A$, the CP matrix depends on the drift only through $\operatorname{Tr}K$. Since
\begin{equation}
\operatorname{Tr}K = 2\gamma(1-x)-\tfrac{1}{2}Z(D),
\end{equation}
the reduced one-mode problem is
\begin{equation}
\text{minimize } Z(D)
\quad\text{subject to}\quad
M(D):=D+i\bigl(\operatorname{Tr}K\bigr)\sigma\succeq 0.
\end{equation}
Equivalently,
\begin{equation}
M(D)=
\begin{pmatrix}
a & c+i\tau\\
c-i\tau & b
\end{pmatrix}\succeq 0,
\qquad
\tau:=2\gamma(1-x)-\tfrac{1}{2}\!\left(\frac{a}{v_q}+\frac{b}{v_p}\right).
\end{equation}
This is a semidefinite program in the real variables $(a,b,c)$.

\subsection{Standard affine SDP form}

Define
\begin{equation}
A_0:=2\gamma(1-x)
\begin{pmatrix}
0&i\\
-i&0
\end{pmatrix},
\end{equation}
\begin{equation}
A_1:=
\begin{pmatrix}
1 & -\dfrac{i}{2v_q}\\[6pt]
\dfrac{i}{2v_q} & 0
\end{pmatrix},
\qquad
A_2:=
\begin{pmatrix}
0 & -\dfrac{i}{2v_p}\\[6pt]
\dfrac{i}{2v_p} & 1
\end{pmatrix},
\qquad
A_3:=
\begin{pmatrix}
0&1\\
1&0
\end{pmatrix}.
\end{equation}
Then $M(D)=A_0+aA_1+bA_2+cA_3$, and the primal SDP is
\begin{equation}
\boxed{
\begin{aligned}
\text{minimize } & \frac{a}{v_q}+\frac{b}{v_p} \\
\text{subject to } & A_0+aA_1+bA_2+cA_3\succeq 0.
\end{aligned}}
\end{equation}

\subsection{Dual SDP}

The dual variable is a Hermitian matrix $Y\succeq 0$. The dual problem is
\begin{equation}
\boxed{
\begin{aligned}
\text{maximize } & -\operatorname{Tr}(A_0Y) \\
\text{subject to } & \operatorname{Tr}(A_1Y)=\frac{1}{v_q},\\
& \operatorname{Tr}(A_2Y)=\frac{1}{v_p},\\
& \operatorname{Tr}(A_3Y)=0,\\
& Y\succeq 0.
\end{aligned}}
\end{equation}
By weak duality, every dual-feasible $Y$ yields a rigorous lower bound on the primal optimum.

\section{Exact dual witnesses, optimal value, and KKT conditions}

Define the Hermitian rank-one matrices
\begin{equation}
W_\pm:=\Gamma_0^{-1}\pm \frac{i}{\nu}\sigma
=
\begin{pmatrix}
1/v_q & \pm i/\nu\\
\mp i/\nu & 1/v_p
\end{pmatrix}.
\end{equation}
These are positive semidefinite because
\begin{equation}
\det W_\pm=\frac{1}{v_qv_p}-\frac{1}{\nu^2}=\frac{1}{\nu^2}-\frac{1}{\nu^2}=0,
\end{equation}
while their traces $\operatorname{Tr}W_\pm = 1/v_q + 1/v_p = 2x/\nu > 0$ (for $\nu\ge 1$, $r\ge 0$, not both zero). Hence $W_\pm\succeq 0$ and each has rank one.

\subsection{Branchwise dual-feasible optimizers}

For the branch $x>1$, define
\begin{equation}
Y_+:=\frac{\nu}{\nu-1}\,W_+.
\end{equation}
A direct calculation gives
\begin{equation}
\operatorname{Tr}(A_1Y_+)=\frac{1}{v_q},
\qquad
\operatorname{Tr}(A_2Y_+)=\frac{1}{v_p},
\qquad
\operatorname{Tr}(A_3Y_+)=0,
\end{equation}
so $Y_+$ is dual feasible. Its dual value is
\begin{equation}
-\operatorname{Tr}(A_0Y_+)
=\frac{4\gamma(x-1)}{\nu-1}.
\end{equation}
Therefore,
\begin{equation}
Z(D)\ge \frac{4\gamma(x-1)}{\nu-1},\qquad x>1.
\end{equation}

For the branch $x\le 1$, define
\begin{equation}
Y_-:=\frac{\nu}{\nu+1}\,W_-.
\end{equation}
Again, $Y_-$ is dual feasible with value
\begin{equation}
-\operatorname{Tr}(A_0Y_-)
=\frac{4\gamma(1-x)}{\nu+1}.
\end{equation}
Therefore,
\begin{equation}
Z(D)\ge \frac{4\gamma(1-x)}{\nu+1},\qquad x\le 1.
\end{equation}
Combining the two branches gives the exact lower bound
\begin{equation}
Z(D)\ge \frac{4\gamma|x-1|}{\nu-\operatorname{sgn}(x-1)}.
\label{eq:lower_bound_1mode}
\end{equation}

\subsection{Equivalent witness-pairing form}

The same computation can be written more compactly as $\operatorname{Tr}(W_\pm M(D))\ge 0$. Since
\begin{equation}
\operatorname{Tr}(W_+M)=Z(D)+\frac{2}{\nu}\operatorname{Tr}K,
\qquad
\operatorname{Tr}(W_-M)=Z(D)-\frac{2}{\nu}\operatorname{Tr}K,
\end{equation}
and $\operatorname{Tr}K=2\gamma(1-x)-\frac{1}{2} Z(D)$, one recovers the same two branch inequalities immediately.

\subsection{KKT conditions and covariance alignment}

Let $Y_\star$ denote the active dual optimizer: $Y_+=\nu W_+/(\nu-1)$ for $x>1$ and $Y_-=\nu W_-/(\nu+1)$ for $x\le 1$. Since the primal is strictly feasible away from the singular pure endpoint, strong duality holds and the Karush--Kuhn--Tucker conditions are:
\begin{enumerate}
\item \textbf{Primal feasibility:} $M(D_\star)\succeq 0$.
\item \textbf{Dual feasibility:} $Y_\star\succeq 0$, with the affine trace constraints above.
\item \textbf{Complementary slackness:} $M(D_\star)Y_\star=0$.
\end{enumerate}
Because $Y_\star$ is rank one and proportional to $W_\pm$, complementary slackness is equivalent to $M(D_\star)u_\pm=0$, where $u_\pm$ is the support vector of $W_\pm$.

Write
\begin{equation}
M(D)=
\begin{pmatrix}
a & c+i\tau\\
c-i\tau & b
\end{pmatrix}.
\end{equation}
The support vector of $W_+$ is $u_+=(\sqrt{v_p},\,-i\sqrt{v_q})^T/\sqrt{v_q+v_p}$ (up to normalization). Imposing $M(D_\star)u_+=0$ and separating real and imaginary parts yields two independent relations:
\begin{equation}
c=0,
\qquad
\frac{a}{v_q}=\frac{b}{v_p}.
\label{eq:KKT_alignment}
\end{equation}
The same relations follow from the $W_-$ branch. Hence the optimizer must lie on the covariance ray
\begin{equation}
D_\star=c_\star\,\Gamma_0.
\end{equation}
Substituting into the objective gives $Z(D_\star)=2c_\star$, so the unique optimal diffusion tensor is
\begin{equation}
\boxed{D_{\mathrm{opt}}=\frac{Z_{\min}}{2}\,\Gamma_0.}
\end{equation}
Thus covariance alignment is not an ansatz: it is forced by the KKT system.

\subsection{Exact optimum}

The primal optimizer $D_\star = (Z_{\min}/2)\Gamma_0$ achieves the dual lower bound~\eqref{eq:lower_bound_1mode} with equality. Therefore the unrestricted one-mode optimum is exactly
\begin{equation}
\boxed{Z_{\min}=\frac{4\gamma|x-1|}{\nu-\operatorname{sgn}(x-1)}.}
\end{equation}
At $x=1$, one has $Z_{\min}=0$ and hence $D_{\mathrm{opt}}=0$. This is the exact zero-cost point of the unrestricted Gaussian reverse problem.

\section{Exact local Gaussian Petz reverse and strict suboptimality}

The continuous-time Gaussian Petz reverse is constructed by applying the Petz recovery map $\mathcal{R}_{\rho,\mathcal{N}}(\cdot) = \rho^{1/2}\mathcal{N}^\dagger(\mathcal{N}(\rho)^{-1/2}(\cdot)\mathcal{N}(\rho)^{-1/2})\rho^{1/2}$~\cite{Petz1988} in the infinitesimal limit of the pure-loss channel. For the one-mode pure-loss generator with reference state $\Gamma_0 = \operatorname{diag}(\nu e^{2r}, \nu e^{-2r})$, the Petz reverse generator has drift and diffusion obtained by evaluating the adjoint channel and the state-dependent sandwich at covariance level. The resulting Petz reverse diffusion entries are
\begin{equation}
D_{\mathrm{Petz},q}=\frac{2\gamma(\nu-e^{2r})^2}{\nu^2-1},
\qquad
D_{\mathrm{Petz},p}=\frac{2\gamma(\nu-e^{-2r})^2}{\nu^2-1}.
\end{equation}
Contracting with $\Gamma_0^{-1}$ gives
\begin{equation}
Z_{\mathrm{Petz}}
=\frac{4\gamma}{\nu^2-1}\bigl[x(\nu^2+1)-2\nu\bigr].
\end{equation}
Subtracting the exact optimum yields
\begin{equation}
Z_{\mathrm{Petz}}-Z_{\min}=
\begin{cases}
\dfrac{4\gamma(\cosh(2r)-1)}{\nu-1}, & x\le 1,\\[8pt]
\dfrac{4\gamma(\cosh(2r)+1)}{\nu+1}, & x>1.
\end{cases}
\end{equation}
These are strictly positive for every $r>0$, with equality only at $r=0$ (unsqueezed thermal state). So the local Gaussian Petz reverse is strictly suboptimal for every squeezed target.

\section{Fixed-gauge multimode additive lower bound}

Let
\begin{equation}
\Gamma_0=S\Gamma_W S^T,
\qquad
\Gamma_W=\bigoplus_{k=1}^N \nu_k I_2,
\end{equation}
be a Williamson decomposition. Transform to this frame:
\begin{equation}
\widetilde D=S^{-1}DS^{-T},
\qquad
\widetilde K=S^{-1}KS.
\end{equation}

Then the cost decomposes as
\begin{equation}
Z(D)=\operatorname{Tr}(\Gamma_W^{-1}\widetilde D)=\sum_{k=1}^N Z_k,
\qquad
Z_k=\nu_k^{-1}\operatorname{Tr}(\widetilde D_{kk}).
\end{equation}
Because $\sigma$ is block diagonal in the Williamson frame, the $(k,k)$ diagonal $2\times 2$ principal block of the transformed CP matrix $\widetilde{M}_{kk} = \widetilde{D}_{kk} + i(\widetilde{K}_{kk}\sigma_1 + \sigma_1\widetilde{K}_{kk}^T)$ depends only on the corresponding diagonal blocks $\widetilde{K}_{kk}$ and $\widetilde{D}_{kk}$. Since $\widetilde{M}\succeq 0$ implies $\widetilde{M}_{kk}\succeq 0$ for every principal submatrix, the one-mode witness argument applies blockwise as a \emph{necessary} condition. Define
\begin{equation}
G:=(S^TS)^{-1},
\qquad
x_k(S):=\frac{\operatorname{Tr}(G_{kk})}{2\nu_k}.
\end{equation}
Then each mode obeys
\begin{equation}
Z_k\ge \frac{4\gamma|x_k(S)-1|}{\nu_k-\operatorname{sgn}(x_k(S)-1)}.
\end{equation}
Summing over $k$ yields the fixed-gauge lower bound
\begin{equation}
Z(D)\ge \sum_{k=1}^N \frac{4\gamma|x_k(S)-1|}{\nu_k-\operatorname{sgn}(x_k(S)-1)}.
\label{eq:fixed_gauge}
\end{equation}

\section{Gauge-invariant upgrade via Schur--Horn majorization}

When symplectic eigenvalues are degenerate, the Williamson frame is not unique. In a degenerate sector of multiplicity $m_\lambda$ with common symplectic eigenvalue $\bar\nu_\lambda$, the residual gauge freedom is
\begin{equation}
S\mapsto SR,
\qquad
R^{(\lambda)}\in \operatorname{Sp}(2m_\lambda,\mathbb{R})\cap O(2m_\lambda,\mathbb{R})\cong U(m_\lambda).
\end{equation}
Let $G^{(\lambda)}$ be the corresponding metric block. Decompose it into commuting and anticommuting parts relative to the local symplectic form $\sigma_\lambda$:
\begin{equation}
G_C=\frac{G-\sigma_\lambda G\sigma_\lambda}{2},
\qquad
G_A=\frac{G+\sigma_\lambda G\sigma_\lambda}{2}.
\end{equation}
The anticommuting part has traceless diagonal $2\times 2$ blocks, so the local trace scalars $x_k(S) = \operatorname{Tr}(G_{kk})/(2\nu_k)$ depend only on $G_C$. Since $G_C$ is symmetric and commutes with the complex structure, it is equivalent to a complex Hermitian matrix $H^{(\lambda)}$. The local anti-squeezing scalars are the diagonal entries of the normalized Hermitian matrix $H^{(\lambda)}/\bar\nu_\lambda$ under the residual unitary conjugation.

Let $x_k^*$ denote the eigenvalues of this normalized Hermitian matrix. By the Schur--Horn theorem, the vector of diagonal entries of a Hermitian matrix is majorized by its eigenvalue vector. Since the one-mode cost function
\begin{equation}
f_\nu(x):=\frac{4\gamma|x-1|}{\nu-\operatorname{sgn}(x-1)}
\end{equation}
is convex (it is piecewise linear with slopes $-4\gamma/(\nu+1)$ for $x<1$ and $4\gamma/(\nu-1)$ for $x>1$, and the right slope exceeds the left slope for $\nu>1$), the Schur convexity theorem gives
\begin{equation}
\sum_k f_{\bar\nu_\lambda}(x_k(S))\le \sum_k f_{\bar\nu_\lambda}(x_k^*)
\end{equation}
for every admissible gauge choice $S$ within the degenerate cluster.

Since the fixed-gauge lower bound~\eqref{eq:fixed_gauge} holds for every admissible frame $S$, taking the supremum over all frames within degenerate clusters yields
\begin{equation}
Z(D) \ge \sup_S \sum_{k=1}^N f_{\nu_k}(x_k(S)) = \sum_{k=1}^N f_{\nu_k}(x_k^*),
\end{equation}
where the equality follows because (i) $\sum f_{\nu_k}(x_k(S)) \le \sum f_{\nu_k}(x_k^*)$ for every $S$ by Schur convexity, and (ii) the eigenvalue vector itself is realized by an admissible gauge (the one that diagonalizes $H^{(\lambda)}$), so the supremum is attained. Therefore the sharp gauge-invariant multimode lower bound is
\begin{equation}
\boxed{Z(D)\ge \sum_{k=1}^N \frac{4\gamma|x_k^*-1|}{\nu_k-\operatorname{sgn}(x_k^*-1)}.}
\label{eq:gauge_inv}
\end{equation}

\section{Global canonical moving Williamson frame and crossing regularity}

We now prove the mathematical statement behind the reduction of the multimode problem through eigenvalue crossings.

Let
\begin{equation}
\Gamma_t=e^{-2\gamma t}\Gamma_0+(1-e^{-2\gamma t})I,
\qquad t>0.
\end{equation}
Since $\Gamma_t$ is strictly positive and real analytic for every $t>0$, define
\begin{equation}
A(t):=\Gamma_t^{1/2}\,\sigma\,\Gamma_t^{1/2}.
\end{equation}
Then $A(t)$ is a real-analytic real skew-symmetric family, so
\begin{equation}
H(t):=iA(t)
\end{equation}
is a real-analytic Hermitian family. Its eigenvalues are the signed symplectic eigenvalues $\pm \nu_k(t)$. By Rellich's theorem~\cite{rellich1969perturbation}, there exists a complete real-analytic orthonormal eigenbasis on the positive branches:
\begin{equation}
H(t)w_k(t)=\nu_k(t)w_k(t),
\qquad w_j(t)^\dagger w_k(t)=\delta_{jk}.
\end{equation}

\subsection{Orthogonality identity that survives crossings}

Take two positive-branch eigenvectors $w_j,w_k$. Since $A^T=-A$, the transpose of $(iA)$ satisfies $(iA)^T = iA^T = -iA = -H$. Therefore
\begin{equation}
w_j^T(iA)w_k = \bigl((iA)^T w_j\bigr)^T w_k = (-iAw_j)^T w_k = -\nu_j w_j^T w_k.
\end{equation}
But also $w_j^T(iA)w_k = \nu_k w_j^T w_k$. Hence
\begin{equation}
(\nu_j+\nu_k)w_j^T w_k=0.
\end{equation}
Because $\nu_j+\nu_k>0$ for all faithful times, this implies
\begin{equation}
w_j^T w_k=0
\end{equation}
for all $j,k$, including at degeneracies $\nu_j=\nu_k$. This is the key identity that kills the crossing singularity.

\subsection{Real analytic Darboux basis}

Write $w_k=(x_k-i y_k)/\sqrt{2}$ with $x_k,y_k\in\mathbb{R}^{2N}$. The Hermitian orthonormality $w_j^\dagger w_k = \delta_{jk}$ and the bilinear orthogonality $w_j^T w_k=0$ together imply
\begin{equation}
x_j^Tx_k=\delta_{jk},
\qquad
y_j^Ty_k=\delta_{jk},
\qquad
x_j^Ty_k=0.
\end{equation}
Moreover the eigenvalue equation implies that $(x_k,y_k)$ is a symplectic pair: $\Gamma_t^{1/2}\sigma\Gamma_t^{1/2}x_k = -\nu_k y_k$ and $\Gamma_t^{1/2}\sigma\Gamma_t^{1/2}y_k = \nu_k x_k$. Therefore the collection $\{x_1,y_1,\dots,x_N,y_N\}$ forms a global real-analytic orthonormal Darboux basis of $\mathbb{R}^{2N}$.

Assemble these vectors into an orthogonal matrix $O(t)\in SO(2N,\mathbb{R})$, so that
\begin{equation}
O(t)^T A(t) O(t)=\bigoplus_{k=1}^N \nu_k(t)\,\sigma_1,
\qquad
\sigma_1:=\begin{pmatrix}0&1\\-1&0\end{pmatrix}.
\end{equation}
Then define
\begin{equation}
\Gamma_W(t):=\bigoplus_{k=1}^N \nu_k(t)I_2,
\qquad
S_c(t):=\Gamma_t^{1/2}\,O(t)\,\Gamma_W(t)^{-1/2}.
\end{equation}
A direct computation gives $S_c(t)\Gamma_W(t)S_c(t)^T=\Gamma_t$ and $S_c(t)\sigma S_c(t)^T=\sigma$, so $S_c(t)$ is a global real-analytic symplectic moving Williamson frame.

\subsection{Canonicality at crossings}

Define
\begin{equation}
W_c:=S_c^{-1}\dot S_c,
\qquad
G:=(S_c^TS_c)^{-1}.
\end{equation}
The pure-loss kinematics imply the exact commuting-block identity
\begin{equation}
(\nu_k-\nu_j)W_{C,jk}=2\gamma G_{C,jk},
\qquad j\neq k,
\label{eq:kinematic}
\end{equation}
where the subscript $C$ denotes the part commuting with the local symplectic block. Since $W_c$ is analytic, $W_{C,jk}$ remains finite at crossings. Therefore if $\nu_j=\nu_k$, one must have $G_{C,jk}=0$.

So the analytic moving frame is automatically canonical in every degenerate cluster, and the canonical anti-squeezing data are realized directly by the moving frame: $x_k(S_c(t))=x_k^*(t)$. Hence crossings cease to be singular in the moving-frame formulation.

\section{Moving-frame transformation law and CP covariance}

Let $\Gamma=S_c\Gamma_W S_c^T$ and $W_c=S_c^{-1}\dot S_c$. Define the intrinsic moving-frame generator by
\begin{equation}
K_{\mathrm{mov}}:=S_c^{-1}KS_c-W_c,
\qquad
D_{\mathrm{mov}}:=S_c^{-1}DS_c^{-T}.
\end{equation}
Equivalently, $K=S_c(K_{\mathrm{mov}}+W_c)S_c^{-1}$ and $D=S_cD_{\mathrm{mov}}S_c^T$. Substituting into the covariance equation gives exactly
\begin{equation}
\dot\Gamma_W=K_{\mathrm{mov}}\Gamma_W+\Gamma_WK_{\mathrm{mov}}^T+D_{\mathrm{mov}}.
\end{equation}
Moreover, because $W_c\in\mathfrak{sp}(2N,\mathbb{R})$,
\begin{equation}
W_c\sigma+\sigma W_c^T=0.
\end{equation}
Therefore the CP matrix transforms by congruence:
\begin{equation}
D+i(K\sigma+\sigma K^T)
=S_c\Bigl[D_{\mathrm{mov}}+i(K_{\mathrm{mov}}\sigma+\sigma K_{\mathrm{mov}}^T)\Bigr]S_c^T.
\end{equation}
So complete positivity in the lab frame is equivalent to complete positivity in the moving frame, with no additional connection term.

\section{Reverse-time source law and intrinsic scalar block optimization}

In the reverse matching convention, $K\Gamma+\Gamma K^T+D=2\gamma(\Gamma-I)$. Passing to the moving frame and taking the trace of the $k$-th diagonal $2\times 2$ block gives
\begin{equation}
\dot \nu_k = 2\gamma\nu_k(1-x_k^*).
\end{equation}

Now consider one intrinsic diagonal block. Write the moving-frame drift block and diffusion block as
\begin{equation}
K_k=
\begin{pmatrix}
a & b\\
-b & a
\end{pmatrix},
\qquad
D_k=cI_2,
\end{equation}
so that the scalar source equation is $2a\nu + c = s$ with $s:=2\gamma\nu(1-x)$. The one-mode CP condition $D_k+i(K_k\sigma+\sigma K_k^T)\succeq 0$ becomes $cI_2 + 2ai\sigma \succeq 0$, with eigenvalues $c\pm 2a$. Feasibility is equivalent to $c\ge |2a|$. The block cost is $Z_k=2c/\nu$.

Using $2a\nu=s-c$, feasibility becomes $c\ge |s-c|/\nu$.

\subsection{Case $s\ge 0$}

The minimal feasible point lies on the binding constraint $c\nu = s-c$, giving $(\nu+1)c=s$. Hence
\begin{equation}
c_{\min}=\frac{s}{\nu+1},
\qquad
Z_{\min}=\frac{2s}{\nu(\nu+1)}.
\end{equation}

\subsection{Case $s\le 0$}

The minimal feasible point lies on the binding constraint $c\nu = c-s$, giving $(\nu-1)c=-s$. Hence
\begin{equation}
c_{\min}=\frac{-s}{\nu-1},
\qquad
Z_{\min}=\frac{-2s}{\nu(\nu-1)}.
\end{equation}
Substituting $s=2\gamma\nu(1-x)$ yields exactly
\begin{equation}
\boxed{Z_{\min}=\frac{4\gamma|x-1|}{\nu-\operatorname{sgn}(x-1)}.}
\end{equation}

\section{Global exact multimode optimizer on faithful intervals}

Choose the intrinsic moving-frame optimizer blockwise by setting all off-diagonal intrinsic blocks to zero: $(D_{\mathrm{mov}}^*)_{jk}=0$ and $(K_{\mathrm{mov}}^*)_{jk}=0$ for $j\neq k$. On each diagonal block choose the exact scalar optimizer above with source $s_k = 2\gamma\nu_k(1-x_k^*)$:
\begin{equation}
(D_{\mathrm{mov}}^*)_{kk}=\frac{\nu_k}{2}\,f_{\nu_k}(x_k^*)\,I_2,
\qquad
f_\nu(x)=\frac{4\gamma|x-1|}{\nu-\operatorname{sgn}(x-1)}.
\end{equation}
Then:
\begin{enumerate}
\item The intrinsic covariance equation holds exactly mode by mode.
\item The intrinsic CP matrix is block diagonal and positive semidefinite by the scalar optimization.
\item The cost decomposes exactly as $\operatorname{Tr}(\Gamma^{-1}D^*)=\sum_{k=1}^N f_{\nu_k}(x_k^*)$.
\item Pushing back to the lab frame by $K^*=S_c(K_{\mathrm{mov}}^*+W_c)S_c^{-1}$ and $D^*=S_cD_{\mathrm{mov}}^*S_c^T$ gives an exact Gaussian CP reverse protocol that is globally continuous on every faithful interval.
\end{enumerate}
This attains the gauge-invariant lower bound~\eqref{eq:gauge_inv} with equality. Therefore, on every faithful interval,
\begin{equation}
\boxed{
\inf Z(D_t)=\sum_{k=1}^N \frac{4\gamma|x_k^*(t)-1|}{\nu_k(t)-\operatorname{sgn}(x_k^*(t)-1)}
}
\end{equation}
and this infimum is attained by a globally continuous Gaussian CP protocol.

\section{Integrated action law}

Let $(K_t,D_t)$ be any exact continuous Gaussian reverse protocol on $[\varepsilon,T]$, $\varepsilon>0$, tracking the pure-loss path $\Gamma_t=e^{-2\gamma t}\Gamma_0+(1-e^{-2\gamma t})I$. Applying the pointwise multimode law at each time and integrating gives
\begin{equation}
\int_\varepsilon^T \operatorname{Tr}(\Gamma_t^{-1}D_t)\,dt
\ge
\sum_k\int_\varepsilon^T
\frac{4\gamma|x_k^*(t)-1|}{\nu_k(t)-\operatorname{sgn}(x_k^*(t)-1)}\,dt.
\end{equation}

\section{Pure-endpoint asymptotics and the universal $2/t$ divergence}

Take a pure squeezed one-mode target $\Gamma_0=\operatorname{diag}(e^{2r},e^{-2r})$ with $r>0$. Along the pure-loss path with $\eta=e^{-2\gamma t}$,
\begin{equation}
\Gamma_t=\operatorname{diag}(\eta e^{2r}+1-\eta,\; \eta e^{-2r}+1-\eta).
\end{equation}
The determinant is
\begin{equation}
\det\Gamma_t
=1+2\eta(1-\eta)(\cosh(2r)-1).
\end{equation}
Since $\eta=1-2\gamma t+O(t^2)$ and $1-\eta=2\gamma t+O(t^2)$,
\begin{equation}
\det\Gamma_t=1+4\gamma t(\cosh(2r)-1)+O(t^2).
\end{equation}
Therefore
\begin{equation}
\nu(t)=\sqrt{\det\Gamma_t}
=1+2\gamma t(\cosh(2r)-1)+O(t^2).
\end{equation}
Also, $x(t)=\operatorname{Tr}(\Gamma_t)/(2\det\Gamma_t)\to \cosh(2r)>1$ as $t\to 0^+$. Substituting into the exact one-mode optimum on the $x>1$ branch,
\begin{equation}
Z_{\min}(t)=\frac{4\gamma(x(t)-1)}{\nu(t)-1}.
\end{equation}
Using $x(t)-1 = \cosh(2r)-1+O(t)$ and $\nu(t)-1 = 2\gamma t(\cosh(2r)-1)+O(t^2)$,
\begin{equation}
Z_{\min}(t)=\frac{4\gamma[\cosh(2r)-1+O(t)]}{2\gamma t[\cosh(2r)-1]+O(t^2)}
=\frac{2}{t}+O(1).
\end{equation}
The squeezing magnitude cancels from the leading term. Thus
\begin{equation}
\boxed{Z_{\min}(t)=\frac{2}{t}+O(1),\qquad t\to 0^+.}
\end{equation}
Integrating, $\int_\varepsilon^T Z_{\min}(t)\,dt = -2\ln\varepsilon+O(1)$, which diverges as $\varepsilon\to 0$. Exact continuous Gaussian reversal to a pure nonclassical endpoint therefore requires infinite covariance-weighted diffusion action.

\section{Faithful one-mode fluctuation-cost divergence}

For faithful times $t>0$, the one-mode fluctuation de Bruijn identity gives
\begin{equation}
\dot S_{\mathrm{fluc,min}}(t)=\frac{\nu(t)\,\ell(\nu(t))}{2}\,Z_{\min}(t),
\qquad
\ell(\nu):=\ln\frac{\nu+1}{\nu-1}.
\end{equation}
Using $\nu(t)=1+2\gamma t(\cosh(2r)-1)+O(t^2)$, we have $\ell(\nu(t))=\ln(1/t)+O(1)$. Together with $Z_{\min}(t)=2/t+O(1)$, this yields
\begin{equation}
\dot S_{\mathrm{fluc,min}}(t)=\frac{\ln(1/t)}{t}+O(1/t).
\end{equation}
Hence $\int_\varepsilon^T \dot S_{\mathrm{fluc,min}}(t)\,dt = \frac{1}{2}[\ln(1/\varepsilon)]^2+O(\ln(1/\varepsilon))$. The fluctuation contribution diverges more strongly than the action itself.

\end{document}